\documentclass{article}
\usepackage{icmc2025_paper_template}
\usepackage{times}
\usepackage{ifpdf}
\usepackage{soul}
\usepackage{amsmath}
\usepackage[english]{babel}

\def\papertitle{Real-time implementation of vibrato transfer as an audio effect}
\def\firstauthor{Jeremy Hyrkas} 

\newif\ifpdf
\ifx\pdfoutput\relax
\else
   \ifcase\pdfoutput
      \pdffalse
   \else
      \pdftrue
  \fi
\fi

\ifpdf 
  \usepackage[pdftex,
    pdftitle={\papertitle},
    pdfauthor={\firstauthor},
    bookmarksnumbered, 
    pdfstartview=XYZ 
   ]{hyperref}

  \usepackage[pdftex]{graphicx}
  \graphicspath{{./figures/}}
  \DeclareGraphicsExtensions{.pdf,.jpeg,.png}

  \usepackage[figure,table]{hypcap}

\else 
  \usepackage[dvips,
    bookmarksnumbered, 
    pdfstartview=XYZ 
  ]{hyperref}  

  \usepackage[dvips]{epsfig,graphicx}
  \graphicspath{{./figures/}}
  \DeclareGraphicsExtensions{.eps}

  \usepackage[figure,table]{hypcap}
\fi

\hypersetup{
    colorlinks,%
    citecolor=black,%
    filecolor=black,%
    linkcolor=black,%
    urlcolor=black
}

\title{\papertitle}

\oneauthor
   {\firstauthor} {University of California San Diego  \\ %
     {\tt \href{mailto:jhyrkas@ucsd.edu}{jhyrkas@ucsd.edu}}}

\begin{document}
\capstartfalse
\maketitle
\capstarttrue
\begin{abstract}
An algorithm for deriving delay functions based on real examples of vibrato was recently introduced and 
can be used to perform a vibrato transfer, in which the vibrato pattern of a target signal is imparted onto
an incoming sound using a delay line.
The algorithm contains methods that computationally restrict a real-time implementation.
Here, a real-time approximation is presented that incorporates an efficient fundamental frequency estimation algorithm and
time-domain polyphase IIR filters that approximate an analytic signal.
The vibrato transfer algorithm is further supplemented with a proposed method to transfer the amplitude modulation of the
target sound, moving this method beyond the capabilities of typical delay-based vibrato effects.
Modifications to the original algorithm for real-time use are detailed here and available as source code for an implementation as a VST plugin.
This algorithm has applications as an audio effect in sound design, sound morphing, and real-time vibrato control of synthesized sounds.
\end{abstract}

\section{Introduction}\label{sec:introduction}
Vibrato is a musical technique in which an otherwise stable note is varied in a quasiperiodic manner.
Vibrato manifests as both amplitude and frequency modulation (AM and FM), with the degree of each modulation depending on both
the musician and instrument.
The modulation patterns of vibrato are highly relevant to musical perception.
Vibrato patterns improve the ability of a listener to count the number of separate instruments in a sound example~\cite{Stoter2013}, and modulation patterns can be used to perform source separation of sources playing in unison~\cite{Stoter2014, Stoter2016}.
Vibrato can impart different AM and FM patterns, and may even present differently in each harmonic~\cite{Zhang2015}.
Nevertheless, the vibrato audio effect is often implemented using a variable delay-line, imparting a perfectly periodic pitch shift to an
incoming signal and no amplitude modulation.
As a result, while the vibrato audio effect has a unique sound, it is often a poor substitute for emulating realistic vibrato.

Previous works have demonstrated methods for controlling the vibrato of a recorded sound.
Commonly, the sound is modeled sinusoidally so that vibrato patterns can be removed or expanded in resynthesis.
One state of the art approach captures the AM and FM patterns of each harmonic of the sound,
as well as the residual noise~\cite{Roebel2011}.
Modulation patterns are captured by modeling the trajectory of each harmonic's frequency and amplitude as a sum of low-passed and residual signals, where
the latter contains the vibrato patterns.
In theory, these residual components could be used in a sinusoidal resynthesis of another sound to transfer the vibrato from one signal to another, although this approach has not been demonstrated.

A method was recently proposed to derive a delay function from a sound example that can be used to transfer vibrato from one signal to another, or to
demodulate the vibrato of the original signal~\cite{Hyrkas2024}.
This method operates entirely on the FM caused by vibrato and does not account for AM.
The proposed method requires extensive analysis and/or a priori knowledge of the analyzed signal, limiting its immediate use as a real-time audio effect.
Here, we approximate the existing method using a real-time pitch estimator, an analytic signal approximation method, and several algorithmic choices to
preserve delay function stability.
We additionally propose a simple method for transferring some AM characteristics of a target signal onto an incoming signal inspired by the lowpass plus residual approach~\cite{Roebel2011}.
Vibrato transfer is implemented as a real-time VST plugin, where one signal is analyzed in real-time to control the amplitude and delay line modulation
of another incoming sound.
The amounts of AM and FM transferred are parameterized as user controls.


\section{Vibrato Transfer: offline method}\label{sec:offline_method}
Here, we briefly review previous vibrato estimation methods that lend themselves to vibrato transfer.
This work largely builds on the estimation of a delay function to perform FM transfer, discussed in Section \ref{sec:offline_fm}.
Section \ref{sec:offline_am} discusses a method to model the per-harmonic AM caused by vibrato when using sinusoidal analysis and resynthesis.
We propose a simplified version of the offline method to transfer AM characteristics at the signal level.

\begin{figure*}
\centering
\includegraphics[scale=0.8]{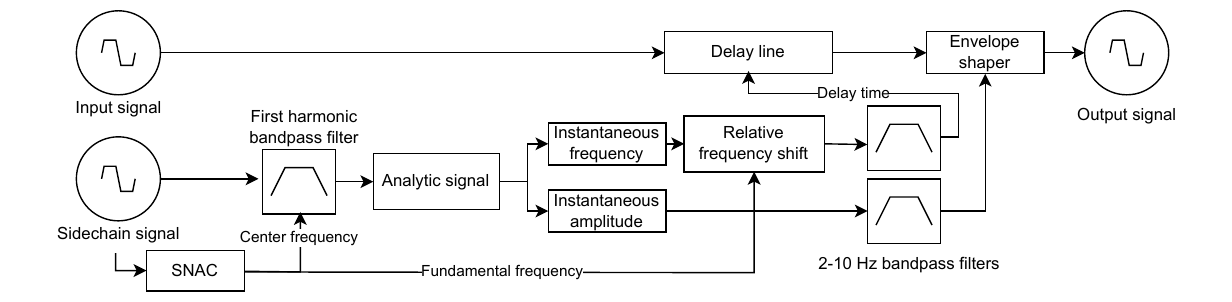}
\caption{The overall signal flow of the real-time vibrato transfer algorithm. A sidechain signal is analyzed and processed to set the delay time and
envelope of the input signal. The FM and AM patterns of vibrato in the sidechain are transferred to the input signal.}
\label{fig:process}    
\end{figure*}

\subsection{Frequency modulation using a variable delay line}\label{sec:offline_fm}
The pitch shift caused by delay-based vibrato effects, sometimes called the \emph{relative frequency shift} or RFS~\cite{Smith2010},
is:
\begin{equation}
  1 - \frac{\omega_i(n)}{\omega_c} = \dot{D}(n),
\label{eq:rfs}
\end{equation}
where $\omega_i(n)$ is the instantaneous frequency of the affected signal, $\omega_c$ is the true sounding frequency, and $\dot{D}(n)$
is the derivative of the delay function $D(n)$.
RFS is the fractional deviation from $w_c$ over time.
The RFS of a signal with real vibrato can be numerically integrated to obtain a $D(n)$ with the same pitch shift pattern when used in a delay line.

Previous work provides two methods for computing \eqref{eq:rfs} from a real signal~\cite{Hyrkas2024}.
In one, the frequency of the first harmonic is tracked using peak-picking across frames of the Short-time Fourier transform (STFT)
of the signal. This method uses a very small hop size (on the order of 8 samples).
$\omega_i(n)$ is the signal of the peak frequency per-frame interpolated to signal length, and
$\omega_c$ set as either the mean or half the addition of the minimum and maximum of $\omega_i(n)$.

The small hop size used in the peak-picking method is computationally expensive and not well suited for a real-time audio effect.
An alternate method centers a bandpass filter around the first harmonic of the signal.
A Butterworth bandpass filter is used, centered around an $f_0$ determined using peak-picking,
with a bandwidth large enough to capture the expected range of natural vibrato.
For a monophonic signal, a bandwidth of half of one octave or less is appropriate.
In monophonic signals, the output of the bandpass filter is expected to be largely sinusoidal, so $\omega_i(n)$ is approximated using
the instantaneous frequency of its analytic signal.

Both methods derive a theoretical delay function that is zero-centered, which may result in a non-causal system when the delay time is negative.
This issue is solved by adjusting the delay function to be purely positive and zero-padding the input signal,
which requires knowledge of the full delay function and cannot be done in real-time.

\subsection{Amplitude modulation using a filtered envelope}\label{sec:offline_am}
Previous work controls the AM caused by vibrato in a sinusoidal analysis and resynthesis of an input signal~\cite{Roebel2011}.
The spectral envelope of the signal is modeled as:
\begin{equation}
    S(\omega,n) = \bar{S}(\omega,n) + \tilde{S}(\omega,n),
    \label{eq:spec_env}
\end{equation}
where $S(\omega,n)$ is the envelope of frequency $\omega$ at time $n$, $\bar{S}$ is the low-passed envelope and $\tilde{S}$ is the residual high-passed
envelope.
$\bar{S}$ is derived using a high order FIR lowpass filter.
Forward-backward filtering is used to form a zero-phase filter.
The residual envelope $\tilde{S}$ contains the AM imparted by vibrato and could be transferred to another signal.
The high order of the filter combined with the need for forward and reverse filtering precludes real-time use.

\section{Real-time approximation}\label{sec:online_method}

The algorithms detailed in Section \ref{sec:offline_method} are adapted here for real-time vibrato transfer.
In this implementation, an input signal is processed by a delay line followed by an envelope shaper.
A sidechain signal is analyzed in parallel, and the delay time and envelope applied to the input are modulated based on the analysis.
If the sidechain signal has vibrato, its FM patterns are transferred using the modulated delay time and its AM patterns by the envelope shaper.
Figure \ref{fig:process} depicts the signal flow of the implementation.

\subsection{Real-time delay function estimation}
\label{sec:online_fm}

Of the two previously proposed methods for delay function estimation~\cite{Hyrkas2024}, deriving $w_i$ in \eqref{eq:rfs} using the
instantaneous frequency of the output of a bandpass filter is preferable to the peak-picking method due to 
the computational load of performing an STFT with a very small hop size. However, choosing the center frequency of the filter and the
denominator $w_c$ in \eqref{eq:rfs} requires some method for $f_0$ estimation.

We use the Specially Normalized AutoCorrelation~\cite{McLeod2008} algorithm (SNAC) due to its efficient computation and accuracy 
when compared against other fast estimators. SNAC uses autocorrelation with a custom normalization
window that is computed with every analysis frame. 
Based on an implementation guide~\cite{VetterHelmholtz}, we compute the normalization window and then bias the autocorrelation slightly using
a triangular window with a slope of $-0.2$ across the length of the analysis signal.
The predicted $f_0$ for each analyzed window is chosen using the period of the highest peak in the autocorrelation signal after normalization and biasing.
The peak is parabolically interpolated, allowing for $f_0$ detection using non-integer periods in samples.
SNAC is computed every 2048 samples, allowing for the detection of an $f_0$ below 50 Hz even when sampling at 96 kHz.
The center frequency of the bandpass filter and $w_c$ in \eqref{eq:rfs} are set based on the $f_0$ estimated using SNAC.

The previously proposed method uses frequency domain processing to derive the analytic signal of the bandpass filter's output. 
Here, we use a previously proposed analytic signal estimator that utilizes two polyphase IIR filters, each containing a cascade of four biquad filters~\cite{VetterHilbert}. 
This estimation eliminates the need to compute the STFT to derive an analytic signal.
Slight inaccuracies in the estimation introduce barely audible jitter when calculating \eqref{eq:rfs}, but are mitigated by 
applying a 2--10 Hz Butterworth bandpass filter to \eqref{eq:rfs} before integrating to obtain the delay function.

Finally, to add further control over the vibrato transfer, we add a user parameter $\alpha_F$, which is used as a scalar for $D(n)$.
Setting $\alpha_F > 1$ imparts a pitch shift on the input that is greater than that of the sidechain. 
Setting $\alpha_F < 1$ reduces the pitch shift; at $\alpha_F=0$, only AM is transferred.

\subsection{Real-time AM estimation}
\label{sec:online_am}

Previous work on vibrato treats time-varying amplitude as the addition of a low-passed envelope
and a high-passed residual that contains the vibrato patterns (see \eqref{eq:spec_env})~\cite{Roebel2011}.
We use a simplified version of this approach.
Instead of a lowpass filter, we use a bandpass filter operating between 2--10 Hz to detect changes in an envelope.
This range is chosen so that the attack (high frequency) and decay or sustain (low frequency) components of the envelope are attenuated and only
those modulations caused by vibrato are retained.
We use a fourth order Butterworth filter instead of the high-order FIR filters used in previous work.

For simplicity, we apply this filter to the instantaneous amplitude of the analytic signal estimated in Section \ref{sec:online_fm}.
This signal contains the envelope of the first harmonic, which may contain different AM patterns than those found in other harmonics~\cite{Zhang2015} or the signal's noise ~\cite{Roebel2011}.
Despite this limitation, we use the first harmonic to avoid the additional computations required to analyze additional harmonics or to compute the RMS of the sidechain.

The AM extracted from the instantaneous amplitude of the first harmonic is used in an envelope shaper,
which adds the modulation to a gain of $0.707$ (-3 dB) so that
\begin{equation}
    y(n) = (0.707 + \alpha_{A} e(n)) \cdot x(n),
\end{equation}
where $x(n)$ is the input to the envelope shaper, $e(n)$ is the envelope containing AM patterns and
$\alpha_{A}$ is a user parameter that can be used to change the depth of modulation.
Setting $\alpha_{A} > 1$ is useful as the depth of AM in the first harmonic is often smaller than that of the full signal.
Setting $\alpha_{A} = 0$ turns off AM transfer and only FM is transferred.

\subsection{Bookkeeping, safeguards and practical details}
\label{sec:bookkeeping}
The delay function derived by integrating \eqref{eq:rfs} is theoretical and zero-centered in a perfect derivation.
Delaying a signal by a negative delay time results in a non-causal system.
Additionally, when $w_c$ is not precisely accurate, \eqref{eq:rfs} can have a small direct current (DC) component, which results in a linearly growing
delay time when integrated~\cite{Hyrkas2024}.
To address these problems in real-time, we use a larger buffer (4096 samples) than typical for a vibrato effect, and operate with an initial
read latency of 512 samples.
In practice, this mitigates the issue of subtracting a negative time from the read pointer and the growth in delay functions for all control signals tested.

To keep the output of the first harmonic bandpass filter stable, we only analyze the sidechain when it is sufficiently loud and the $f_0$ is stable.
When the volume of the sidechain is below -60 dBFS or when the estimated $f_0$ deviates from previous frames, the sidechain analysis stops controlling
the delay time and the envelope shaper. 
When analysis is turned off, the delay time is returned to 512 samples and the envelope is reset; both occur using a slow return rate to avoid noticeable pitch changes and pops.

\section{Implementation, results and evaluation}\label{sec:evaluation}

\begin{figure}
    \centering
    \includegraphics[width=\columnwidth]{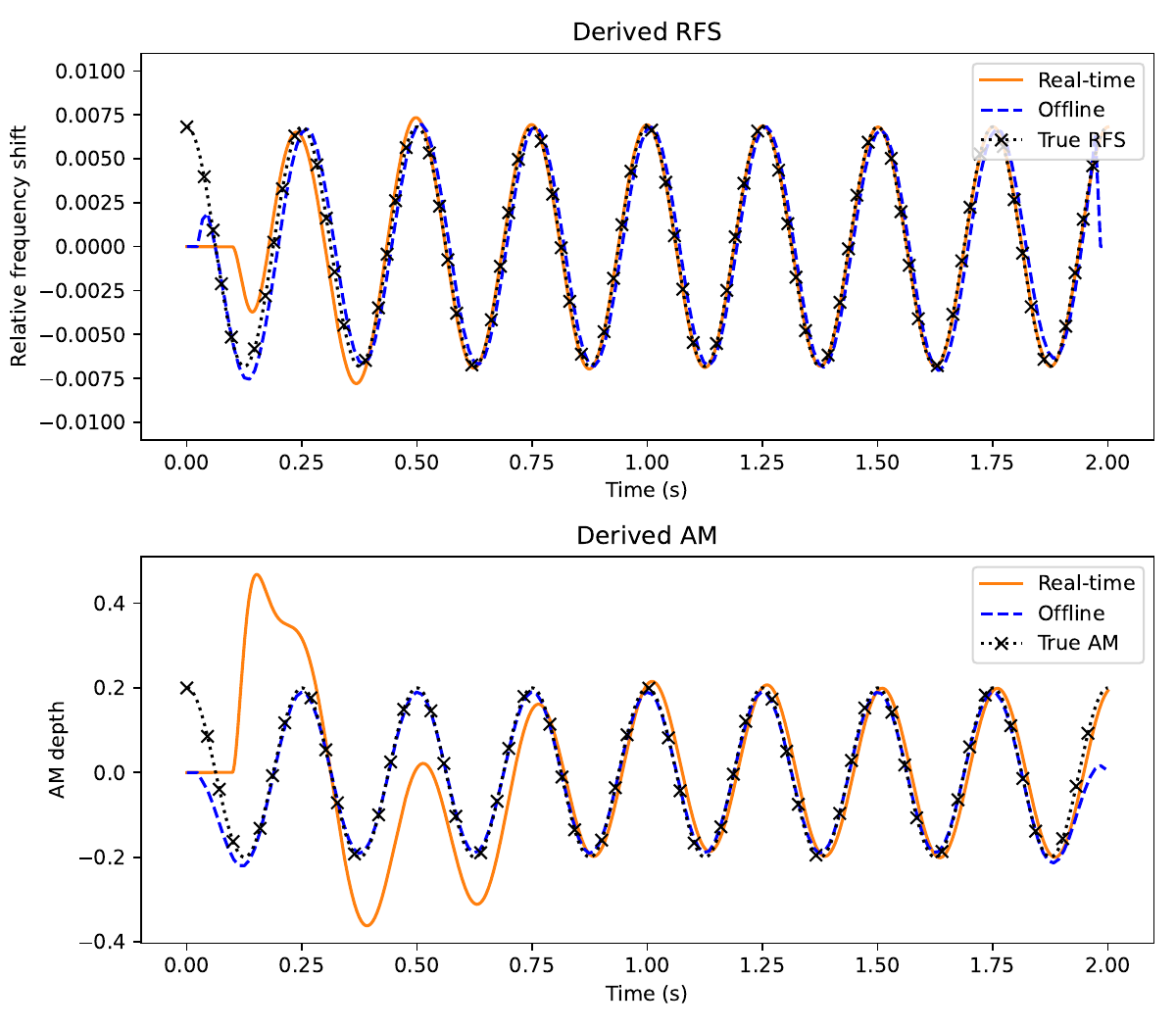}
    \caption{The RFS (top) and AM (bottom) of a modulated sinusoid.
    The real-time algorithm approximates the true signals and offline estimations,
    outside of issues at the onset and a phase distortion of $\sim$ 10--20 ms.}
    \label{fig:sin_example}
\end{figure}

A prototype of the real-time algorithm was implemented in Python and tested against offline baseline methods to ensure that
the streaming version approximates the results of previous work.
Figure \ref{fig:sin_example} compares the RFS and AM envelope recovered from the offline baselines and the real-time implementation
on a modulated sinusoid, where the true RFS and AM are known.
The latency time of the real-time algorithm (i.e. the period at the beginning of each estimation where the RFS and AM signals are 0) is a
result of the safeguards described in Section \ref{sec:bookkeeping};
as a heuristic, we require at least four consecutive approximately equal $f_0$ analyses before the vibrato processing begins.

\begin{figure}
    \centering
    \includegraphics[width=\columnwidth]{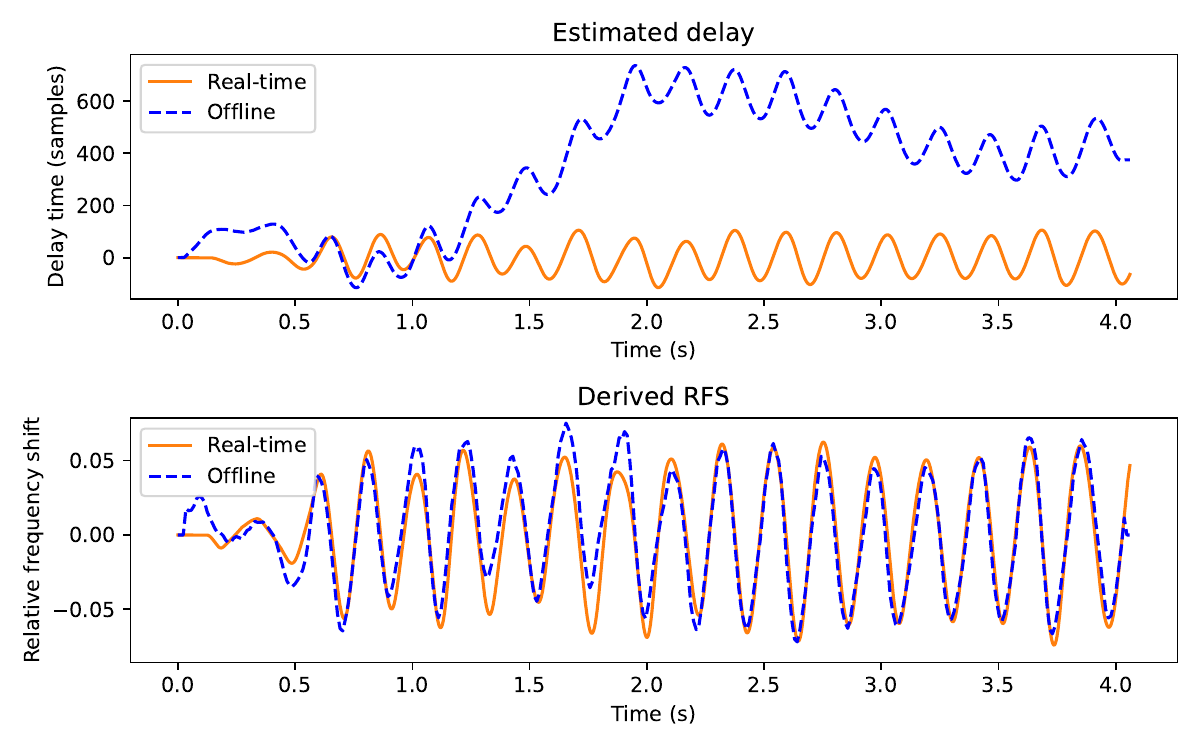}
    \caption{The estimated delay function (top) and RFS (bottom) using the offline baseline and real-time approximation on a recording of a vocalist.}
    \label{fig:voice_delay}
\end{figure}

The real-time approximation of the RFS adheres closely to the true RFS and the baseline method. 
The real-time RFS signal in Figure \ref{fig:sin_example} is slightly out of phase with the true RFS and baseline method by an order of roughly 10 ms due to the bandpass filter on \eqref{eq:rfs} (see Figure~\ref{fig:process}).
Figure \ref{fig:voice_delay} shows the derived delay functions and RFS signals using a real vocal signal.
The offline method suffers from a delay drift (a known issue~\cite{Hyrkas2024}), which the real-time signal avoids due to the bandpass filter's rejection of DC.
The RFS signals are comparable in contour and amplitude, with slight deviations that are acceptable given the real-time method's more stable delay function.

The proposed AM transfer method is a reasonable approximation of the true AM but has two notable deficiencies.
First, a signal onset causes the shape of the AM to be exaggerated while the IIR filter stabilizes.
Second, the resulting AM signal is offset from the true AM and the baseline method by around 20 ms (this range was observed in
multiple test signals).
The baseline method uses an FIR filter with over 10,000 taps and requires forward-backward filtering to recover
the original phase~\cite{Roebel2011}.
Given that the baseline is impossible in real-time, the issues observed in the AM approximation are tolerable for
the purpose of vibrato transfer as an audio effect.
However, future work could focus on more accurate real-time AM estimation.

\begin{figure}
\centering
\includegraphics[width=\columnwidth]{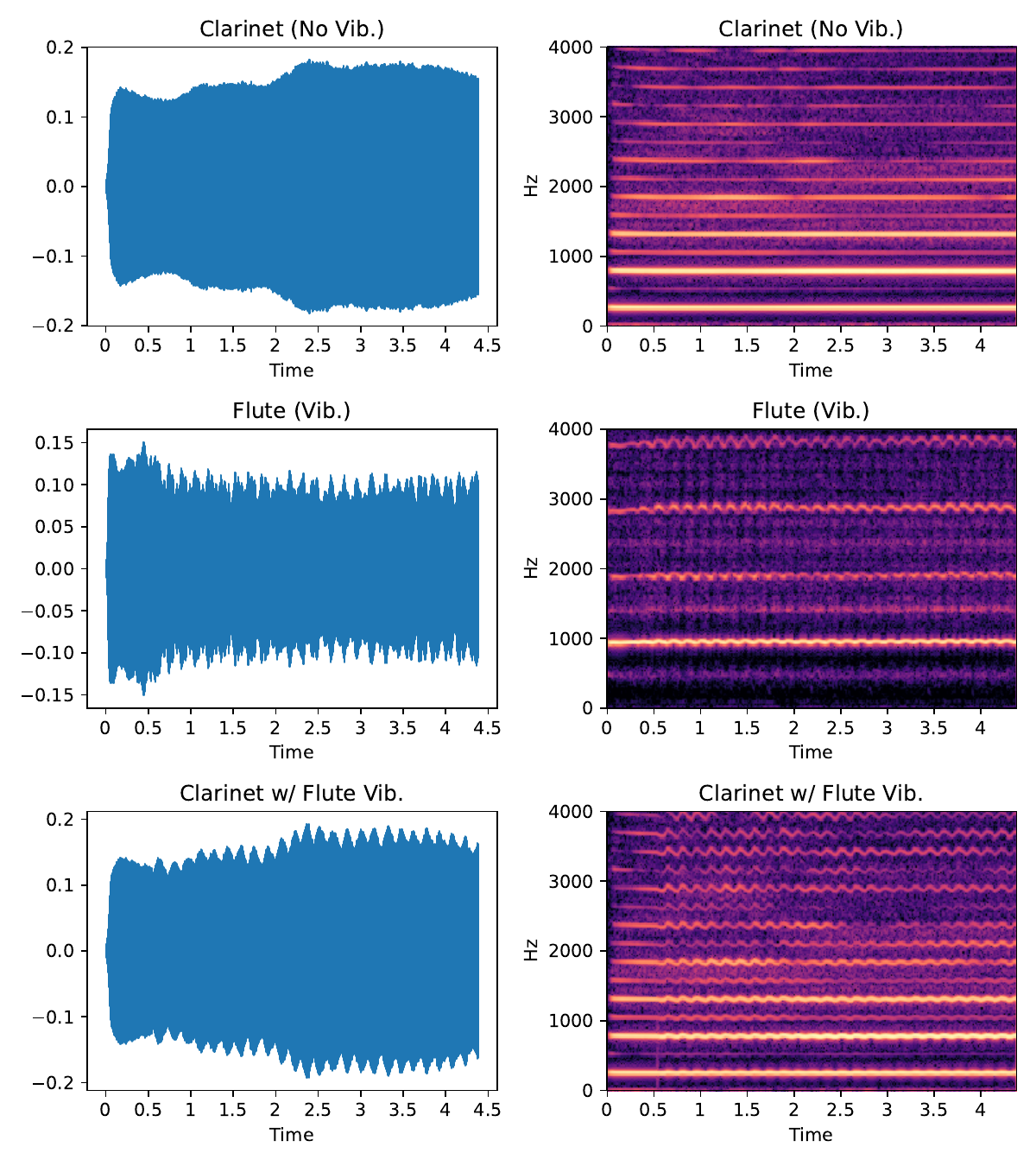}
\caption{Vibrato patterns are transferred from a flute (middle) to a clarinet (top: pre-transfer, bottom: post-transfer).
AM is more visible in the waveforms (left) with FM more visible in the spectrograms (right).}
\label{fig:transfer}
\end{figure}

A proof-of-concept implementation of real-time vibrato transfer is provided as a VST plugin using the JUCE framework.
The waveforms and spectrograms of a vibrato transfer using this implementation are shown in Figure \ref{fig:transfer}, where the vibrato patterns
from a flute signal are transferred to a clarinet.
The code and more sound examples demonstrating real-time vibrato transfer are available online
\footnote{\url{https://github.com/jhyrkas/VibratoTransfer}}.


\section{Discussion}\label{sec:discussion}

Real-time vibrato transfer has multiple applications in music making and sound design.
The vibrato audio effect, a popular tool for guitar and synthesizer, can be supplemented with
real-time control from an external signal to create a more expressive and realistic vibrato.
The incorporation of amplitude modulation supplements the typical delay-line implementation to impart a vibrato closer in timbre to acoustic instruments.
Transferring vibrato from recordings of acoustic instruments may be used to complement sounds synthesized using physical modeling, which may not
account for vibrato.
Finally, since distinct vibrato patterns are perceptually relevant to source separation~\cite{Stoter2014}, matching vibrato patterns of unison signals offers a unique approach to blending signals in sound design.

Previous algorithms to model vibrato in a transferable way~\cite{Roebel2011,Hyrkas2024} cannot be implemented in real-time.
This work approximates previous results in real-time by using fast pitch detection, analytic signal approximation and safeguards against issues arising from a theoretical delay estimation and practical limitations.
The AM transfer method proposed here is acceptable as a proof-of-concept, but issues in the onset and phase delay of the estimated AM can be improved in future work.
A real-time implementation in C++ is available for use and can be further optimized in open source or commercial adaptations.

\begin{acknowledgments}
The methods detailed here were developed under advisement from Tamara Smyth and Tom Erbe at UCSD.
\end{acknowledgments} 

\bibliography{icmc_vibrato}

\end{document}